\documentclass[prb,twocolumn,showpacs,floatfix]{revtex4}
\usepackage{dcolumn,epsf}
\usepackage{amsmath}

\begin{document}

\title{KMC-MD Investigations of Hyperthermal Copper Deposition on Cu(111)}

\author{J.M. Pomeroy}
\affiliation{Cornell Center for Materials Research, Cornell
University, Ithaca, NY 14853 U.S.A} 
\email[Email: ]{jmp42@cornell.edu}
\author{J. Jacobsen}
\affiliation{Haldor Tops{\o}e A/S, Denmark}
\author{C.C. Hill}
\affiliation{Gene Network Sciences, Ithaca, NY 14850}
\author{B.H. Cooper\footnote{deceased}}
\affiliation{Cornell Center for Materials Research, Cornell
University, Ithaca, NY 14853 U.S.A} 
\author{J.P. Sethna} 
\affiliation{Cornell Center for Materials Research, Cornell
University, Ithaca, NY 14853 U.S.A} 

\date{\today} 

\begin{abstract}

Detailed KMC-MD (kinetic Monte Carlo-molecular dynamics) simulations
of hyperthermal energy (10-100 eV) copper homoepitaxy have revealed a
re-entrant layer-by-layer growth mode at low temperatures (50K) and
reasonable fluxes (1 ML/s).  This growth mode is the result of atoms
with hyperthermal kinetic energies becoming inserted into islands when
the impact site is near a step edge.  The yield for atomic insertion
as calculated with molecular dynamics near (111) step edges reaches a
maxima near 18 eV.  KMC-MD simulations of growing films find a minima
in the RMS roughness as a function of energy near 25 eV.  We find that
the RMS roughness saturates just beyond 0.5 ML of coverage in films
grown with energies greater than 25 eV due to the onset of
adatom-vacancy formation near 20 eV.  Adatom-vacancy pairs increase
the island nuclei density and the step edge density, which increases
the number of sites available to insert atoms.  Smoothest growth in
this regime is achieved by maximizing island and step edge densities,
which consequently reverses the traditional roles of temperature and
flux: low temperatures and high fluxes produce the smoothest surfaces
in these films.  Dramatic increases in island densities are found to
persist at room temperature, where island densities increase an order
of magnitude from 20 to 150 eV.
\end{abstract}

\maketitle

\section{Introduction}
\label{sec:intro}

Recent technological advances have resulted in wide-ranging
implementation of devices which utilize hyperthermal energy particles
for thin-film and nano-scale device growth\cite{Minv}.  While these
advances have propelled production efforts, a detailed understanding
of the relevant physical mechanisms has not been fully developed.  It
has become increasingly apparent that even hyperthermal energetic
particles can stimulate a variety of thermal and non-thermal
processes, ranging from strain relaxation\cite{Karr} to sputter
erosion\cite{Mich2}, enhanced nuclei densities\cite{Esch}, and
improved composite layer adhesion\cite{Flad1}.

Efforts to model crystal surfaces during deposition have provided
detailed information about the nature of surface kinetics.  For
example, molecular dynamics has allowed accurate predictions of many
energy barriers for surface diffusion processes\cite{Stolze}.  Many of
these predictions have been carefully addressed
experimentally\cite{Morg1,Rosen3,Geis1}, but most efforts use
deposition techniques with thermally generated constituents arriving
at the substrate with less than a tenth of an eV.

Many of these studies have identified the ``Ehrlich-Schw\"oebel''
\cite{ehrl1,schw1} barrier for interlayer diffusion as the source
of three dimensional growth modes.  The resulting ``uphill'' current
produced by this interlayer diffusion barrier can be reduced by
increasing temperature to surmount this barrier\cite{Rosen1,Wulf1} or
sufficiently decreasing step-step separation\cite{geis2}.  However, in
hetero-structures and nano-structures, elevated temperatures result in
inter-diffusion, chemical reaction, and thermodynamic relaxation,
making nano-scale patterns difficult to retain.

Physical vapor deposition (PVD) techniques utilizing hyperthermal
energy constituents produce smoother epitaxial films in many systems
with finite Ehrlich-Schw\"oebel barriers\cite{Esch2,Got1}.  It has been
proposed that a ``peening'' effect known for building-up stress at
medium energies\cite{Sun,Wind1,Wind2} may act as a relaxation
mechanism at low and hyperthermal energies\cite{Koster}.

Progress toward understanding hyperthermal energy collisions has been
hindered by a lack of models which both accurately describe the
collision process and the kinetic processes at realistic deposition
rates ($\sim 1$ ML/s).  The KMC-MD method allows the complexity of the
atomic collision to be modeled uniquely with molecular dynamics for
each atom without prior bias.  Between deposition events, the kinetic
Monte Carlo evolves the system using well understood kinetics, until
the next deposition event.  Previous KMC-MD studies of platinum and
silver were able to provided new insights into the role of the
hyperthermal atom collision during growth\cite{JJ}.

This paper presents results for energetic collisions on the Cu(111)
surface: first, isolated molecular dynamics studies of atom impacts,
and, second, KMC-MD simulations of entire films grown with
hyperthermal energy atoms.  A brief description of the simulation,
improvements, and the energy barriers used for our simulations is
presented first.  The yields for various atomistic mechanisms
resulting from molecular dynamics simulations of isolated impacts is
presented next.  The results of the full KMC-MD growth model at 50 K
are then discussed.  Finally, simulations of sub-monolayer films at
room temperature are presented.

\section{Simulation Details}
\label{sec:simdet}

\begin{table}
\begin{center}
\caption[]{Effective medium (EMT) energy barriers in meV used in the
KMC simulations have been calculated using ARTwork\cite{artwork}.  The
details of the moves are discussed in detail elsewhere\cite{JJ2}. Edge
diffusion are for atoms moving along the edge of an island.  The
``Step'' denotes whether the atom is moving along a (100) or (111)
micro-facet, A step or B step respectively.  $N_i$ indicates the
initial number of in-plane nearest neighbors, and $N_f$ the final.}

\label{tab:energies}
\begin{tabular}{lcccr}
\\
{\bf Terrace Diffusion} & & & & \hspace{0.2in} meV \\ 
Adatom diffusion & & & & 54\\
Diffusion away from a step & & & & 525  \\
Diffusion of dimers & & & & 117 \\
Diffusion of vacancies & & & & 618 \\
Dissociation from 1 NN & & & & 318\\
{\bf Edge Diffusion} & $N_i$ & \hspace{0.2in} Step \hspace{0.2in} &
$N_f$ & \\
Corner of ``A'' island & 1 & B & 1& 179 \\
Corner of ``B'' island & 1 & A & 1 & 60 \\
Corner diffusion & 1 & A & $\ge$ 1 & 44 \\ 
Corner diffusion & 1 & B & $\ge$ 1 & 108 \\
Step to Corner & 2 & A & 1 &  271 \\
Step to Corner & 2 & B & 1 & 351 \\
Step diffusion & 2 & A & $>1$ & 228 \\
Step diffusion & 2 & B & $>1$ & 329 \\
Kink to corner & 3 & A & 1 & 496 \\
Kink to corner & 3 & B & 1 & 580 \\
Kink to step & 3 & A & $>1$ & 436 \\
Kink to step & 3 & B & $>1$ & 525 \\
{\bf Interlayer diffusion} & & & & \\ 
Descent at straight step & & & & 167 \\
Descent at B step kink & & & & 229 \\
\\
\end{tabular}
\end{center}

\end{table}

\begin{table}
\begin{center}
\caption[Energy barriers for vacancy diffusion]{Additional effective
medium energy barriers in meV used in the KMC-MD needed for 273 K
  simulations are presented.  Energy barriers presented here are moves
  from highly coordinated sites, which were not included in low
  temperature simulations since the rates for these moves are
  negligible at low temperatures.}

\label{tab:vacenergies}
\begin{tabular}{llrc}
\\ \hline \hline 
& & & B Step \\ 
Start \hspace{0.5in} & End \hspace{0.5in} & meV & (if different) \\ \hline \\
{\bf 5NN} & 5 NN &  606 \\
{5NN} & 4 NN &  644 \\
{5NN} & 3 NN &  692 \\
{5NN} & 2 NN &  748 \\
\\
{\bf 4NN}& 5NN& 448 \\
{4NN} & 4 NN &  470 \\
{4NN} & 3 NN &  502 \\
{4NN} & 2 NN &  695 \\
4NN & 1 NN & 681 \\
\\
{\bf 3NN} & 5 NN &  214 \\
{3NN} & 4 NN &  291 \\
{3NN} & 3 NN &  322 \\
{3NN} & 2 NN &  436 & 525\\
{3NN} & 1 NN &  496 & 580\\
3 NN & 0 NN & 748 \\
\\
{\bf 2NN} & 5 NN &  127 \\
{2NN} & 4 NN &  $\sim$300 \\
{2NN} & 3 NN &  $\sim$300 \\
{2NN} & 2 NN & 228 & 329 \\
{2NN} & 1 NN & 271 & 351 \\
\hline \hline
\end{tabular}
\end{center}
\end{table}

Accurate modeling of crystal growth with hyperthermal energy atoms
requires accurately modeling two classes of events active at times
scales separated by about six orders of magnitude.  The hyperthermal
atom impact and subsequent thermalization process is complete in about
four picoseconds.  On the other hand, the surface kinetics are active
on the microsecond time scale.  The technical challenge of accurate
modeling sub-picosecond events for microseconds has been resolved by
coupling two techniques, one appropriate for each time scale.
Molecular dynamics is used to simulate the hyperthermal atom
collisions, but is not feasible for modeling surface diffusion at
realistic deposition rates.  For surface diffusion we use kinetic Monte
Carlo, which passes atomic configurations at selected impact sites to
the molecular dynamics, and accepts the new configurations following
an impact.  The full details of these simulations have been detailed
elsewhere \cite{JJ}.
 
We use a bowl-shaped molecular dynamics cluster with three classes of
atoms: fully dynamic atoms nearest to the site of the impact,
surrounded by three layers of dynamic Langevin atoms, and finally, an
outer shell with four layers of static atoms.  The Langevin
coefficients and cell size are tuned to dissipate energy which can
reflect from the boundary and lead to unrealistic rates for energetic
reflection and adatom/vacancy formation.

We have found it useful to model isolated hyperthermal atom collisions
with a few selected atomic configurations to develop a general picture
of which atomic mechanisms are important at various energies.  Many
atomic configurations on the surface can be classified according to
the distance to a step edge, so we have selected several positions
near a (111) step edge.  We model 100 collisions in each of five
atomic cells above a step edge, at the step edge, and five cells below
the step edge for each energy of interest.  For each collision, impact
parameters are randomly selected, and the cluster evolved until the
Langevin atoms have thermalized the system.  The configuration is then
frozen into a final state for analysis and saved.  Once satisfactory
statistics are developed at one position, the impact site is moved one
atomic cell, and the process is repeated.  We have found that the
statistics for the fifth cell above or below is representative of all
cells further from the step.

The role of the molecular dynamics during KMC-MD is the same, except
that the configuration of atoms in the molecular dynamics cell is
determined by the local environment of the impact site.  The KMC uses
a hexagonal lattice with in-plane periodic boundary conditions and a
compliment of 23 pre-defined thermal moves.  Interstitial and HCP
lattice positions are not allowed.  The activation energies for these
moves were calculated using ARTwork\cite{artwork} and are listed in
Table \ref{tab:energies}.  (For the 273 K simulations the barriers in
Table \ref{tab:vacenergies} are included as well.)  Included in the
KMC rate table is a flux weighted choice for adding new atoms.  When
the algorithm chooses to introduce a new atom, an impact site is
randomly selected.  The local configuration of atoms is then copied
into the molecular dynamics cluster, which simulates the collision.
Once the final configuration is determined, it is returned to the KMC,
and thermal evolution continues.

\subsection{Modifications to the KMC-MD algorithm}

One of the trickier parts of the KMC-MD method involves moving atoms
from the continuous MD space to the discrete lattice of the KMC.
During collisions involving many atoms on the surface, clusters of
atoms occasionally freeze into HCP rather than FCC lattice positions.
(HCP and FCC lattice positions are energetically equivalent on the
(111) face using the EMT potential for copper.)  In the previous
studies of platinum and silver\cite{JJ}, these clusters rarely
exceeded five atoms.  As the algorithm encountered atoms in HCP sites
it would place them on the nearest available FCC site.  During
simulations of copper, these clusters are sometimes as large as
eight atoms, and this previous technique did not always preserve the
shape of the cluster.  In some cases, atoms near the middle of a
cluster could not be placed at all, since all nearby sites would
already be filled.

Our modified algorithm creates a list of atoms in HCP sites, then
sorts the list from highest coordination to lowest.  As each atom in
the list is selected for placement, the three fcc sites surrounding
that HCP site are checked for occupancy.  The wayward atom is then
placed into the unoccupied fcc site with the highest coordination (a
random selection occurs if multiple sites have the highest
coordination.)  Since atoms near the center of a cluster get placed
first, all atoms have an available FCC position.  This change
preserves the cluster and has successfully placed all the atoms in
supported FCC sites.

\section{Results for MD Collisions}
\label{sec:mdresults}

\begin{figure}[t]
\begin{center}
\epsfbox{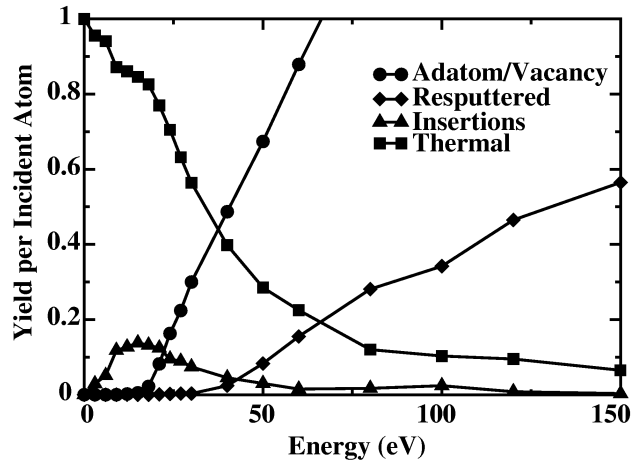}
\end{center}
\caption[]{Molecular dynamics simulations reveal a hierarchy of
energetically activated non-equilibrium events, described in
Sect. \ref{sec:mdresults}.  In order of increasing energy, the
insertion mechanism is activated as low as 3eV, followed by
adatom-vacancy pair formation near 20 eV, and atomic re-sputtering
near 40 eV.}
\label{fig:mdresults}
\end{figure}

Simulating individual atomic collisions in a pre-selected environment
can provide a general insight into the yields for atomic mechanisms at
different energies.  Once an impact has been simulated, the final
atomic configuration is classified according to the change in the
population of the atomic layers.  If the impact site is above the step
edge, and the atom incident atom is incorporated into the step, the
event is considered an insertion.  We do not distinguish between an
atom that is actually inserted, and one which just bounces over the
step edge.  A decrease in the population of a layer requires the
formation of a vacancy.  The formation of vacancies usually provides
additional adatoms (adatom-vacancy pairs) that can contribute to
surface relaxation through enhanced lateral diffusion.  If the total
number of atoms in the cluster decreases, this is considered a
sputtering event.  (Spontaneous thermodynamic re-evaporation is
negligible.)

The yields averaged from many simulated collisions at several
different atomic configurations are presented in Figure
\ref{fig:mdresults}.  The hyperthermal atomic mechanisms observed in
platinum and silver are present in copper, but the specific energies
of activation varying somewhat.  The insertion mechanism is active at
the first position above the step edge as low as 3 eV.  As the energy
increases, atoms are inserted deeper into the island, which increases
the yield.  By 9 eV insertion events are observed four lattice
positions into an island.  The fifth atomic position into the step is
not susceptible to insertion, so the insertion probability reaches a
maxima at 15 eV.  Above this energy, insertions continue to dominate
beyond the first position above the step, but positions near the step
become unstable and often form adatom-vacancy pairs.

Adatom-vacancy formation on the flat terrace begins abruptly near 20
eV, and the total yield increases at a rapid rate, reaching a yield of
one by 60 eV.  On average, more than two adatom-vacancy pairs are
created per incident atom by 150 eV.  By 80 eV, adatom-vacancy has a
higher probability at all the step positions considered than any other
mechanism.

At about 40 eV we begin to observe atoms escaping from the system,
with some preference for positions close to the atomic step edge.  At
low energies, an atom incident just above a step could shift the
registry of atoms and become incorporated, but at higher energies the
transverse momentum provided by the incident atom can eject step atoms
from the system.  The rapid increase in yields for adatom-vacancy
pairs and sputtered atoms combine to double the number of dislodged
atoms between 60 an 100 eV, greatly increasing the surface mobility
and reducing the net growth rate to 65\% of the incident flux.  More
comprehensive studies of re-sputtering have been reported
elsewhere\cite{Wad2}.

\section{Results and Discussion of 50K KMC-MD deposition}
\label{sec:kmcmdres}

\begin{figure}[t]
\begin{center}
\epsfbox{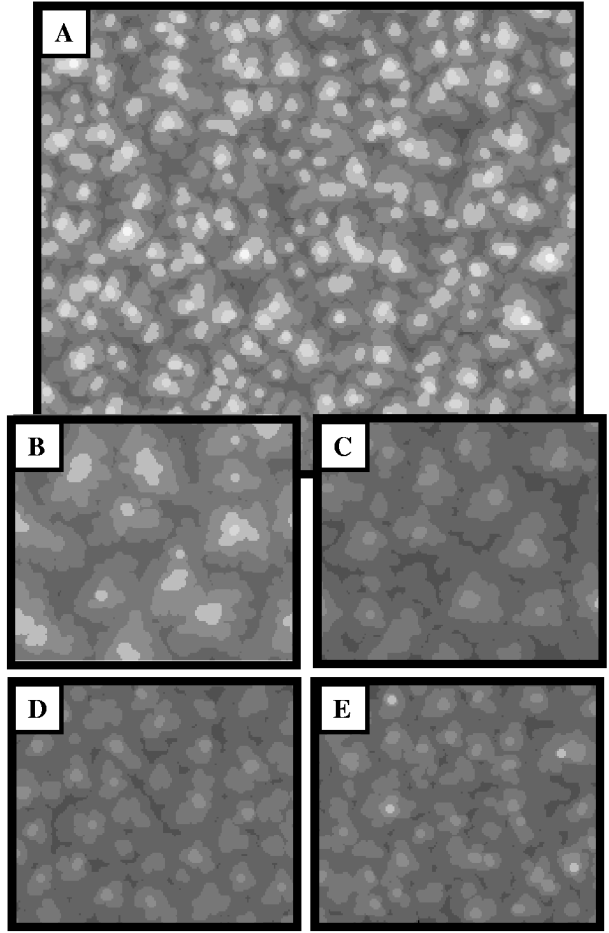}
\end{center}
\caption[]{Gray-scale images for KMC thermal deposition and four
KMC-MD hyperthermal energy depositions are shown: (A) thermal, (B) 12
eV, (C) 21 eV, (D) 30 eV, and (E) 40 eV.  All images have the same
lateral length scales (atom size) and color-maps.  Films were grown at
1 ML/s on an 80x80 lattice (thermal deposition used a 150x150 lattice)
using all the diffusion moves listed in Table \ref{tab:energies}.}
\label{fig:films}
\end{figure}

While the molecular dynamics simulations of isolated impacts estimates
the relative yields of different atomic mechanisms at various
energies, the effect on a dynamically growing film requires the more
sophisticated KMC-MD simulation.  We have used KMC-MD to grow copper
thin films on a Cu(111) surface using energies ranging from thermal to
40 eV.  All the mechanisms identified with the isolated molecular
dynamics in the previous section are active, but the yield for
re-sputtering below 40 eV is negligible. 

Five examples of copper thin films grown with the KMC-MD are presented
in Figure \ref{fig:films}.  The films shown were grown with a wide
range of energies: thermal (A), 12 eV (B), 21 eV (C), 30 eV (D), and
40 eV (E).  All films described in this section are grown at
$\sim$50$^\circ$ K at 1 monolayer/s deposition rate on an 80x80
lattice, except thermal deposition, which had a 150x150 lattice.
System sizes are selected to avoid finite size effects, but all the
images have the same lateral length scale, and the same color map for
ease of comparison (the size of an atom is the same in all the images,
and the layer depths have the same color sequence in all images).
While four mono-layers of copper has been deposited in all cases, the
films grown at 21 eV and 30 eV do not have any atoms in the seventh
and eighth layers.  The thermally deposited film has a large
population of atoms in these upper two layers and is rougher than
those grown with energetic deposition.  The step density in the
thermal films is much higher at 0.74 than the 21 eV or 30 eV films
(0.39 and 0.45, respectively), corresponding to a shorter lateral
length scale (step densities are discussed in more detail later).

\begin{figure}[t]
\begin{center}
\epsfbox{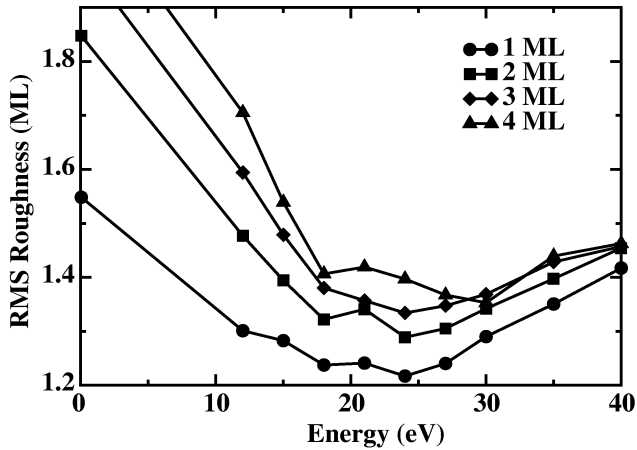}
\end{center}
\caption[]{The RMS roughness at the completion of each monolayer shows
a minima near 25 eV.  As the film grows thicker, the minima is
observed to shift toward higher energies.  Beyond 1 ML of coverage,
the roughness of films grown with energies above 25 eV increases
slower than films grown with lower energies.  The time evolution of
the RMS roughness is shown in greater detail in Figure
\ref{fig:roughvtime}.}
\label{fig:roughvenergy}
\end{figure}

A common way of representing smooth growth that establishes a
connection with experimental efforts\cite{Poels,Mich1} is to plot the
simulated anti-phase Bragg intensities associated with refelection
high energy electron diffraction or x-ray diffraction.  Anti-phase
intensities will exhibit complete oscillations between 0 and 1 for
perfect layer-by-layer growth, and a monotonic decay for
three-dimensional roughening.  Due to space constraints, we have
limited the presentation of simulated anti-phase intensity to our
discussion of temperature and flux in Section \ref{sec:tempflux}.
With the exception of thermally deposited films, all the films studied
exhibited layer-by-layer oscillations of varying strengths.  The
anti-phase intensity of the thermally grown film decays monotonically
in this low-temperature regime, consistent with experimental
observation \cite{Rosen2}.  The anti-phase intensity oscillations are
strongest between 20-30 eV, corresponding with the minima in roughness
shown in Fig
\ref{fig:roughvenergy}.

\begin{figure}[t]
\begin{center}
\epsfbox{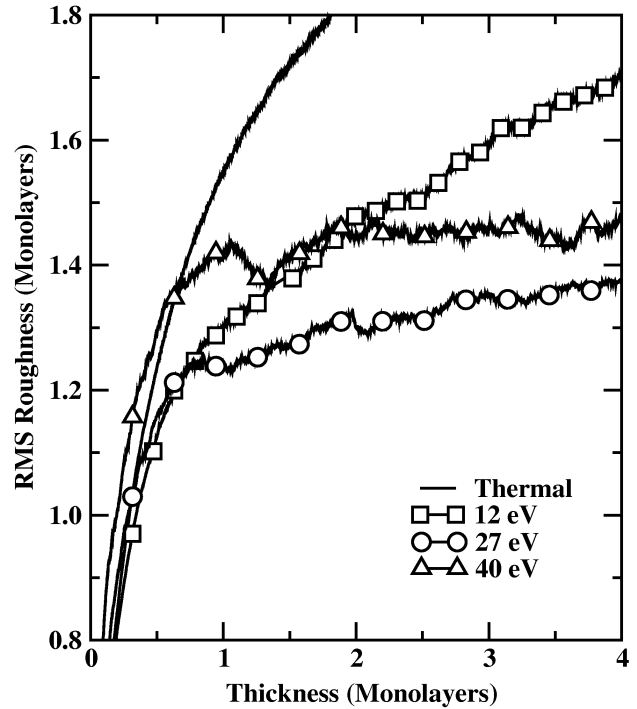}
\end{center}
\caption[]{RMS roughness as a function of time is presented for four
characteristic energies.  During thermal deposition, the RMS roughness
increases steadily over the entire range studied.  At hyperthermal
energies, the roughness is observed to grow more slowly after 0.5 ML.
For films grown with 27 and 40 eV particles, the roughness grows very
slowly above 1 ML compared to thermal deposition. }
\label{fig:roughvtime}
\end{figure}

The roughness of the KMC-MD films is quantified by calculating the RMS
roughness at the completion of each of the four mono-layers deposited.
This roughness data as a function of the deposition energy is shown in
Figure \ref{fig:roughvenergy}.  Even after depositing only one
monolayer, the films grown with atoms in the 20 eV range have a much
lower roughness than those grown with higher or lower energies.  As
the film progresses, this minimum roughness appears to shift toward
higher energies.  The roughness of films grown with less than 25 eV
grows more quickly after 1 monolayer of coverage than the roughness of
the films grown with energies greater than 25 eV.  Careful examination
of the 40 eV data reveals very little change in the surface
roughness after the first monolayer.

\begin{figure}[t!]
\begin{center}
\epsfbox{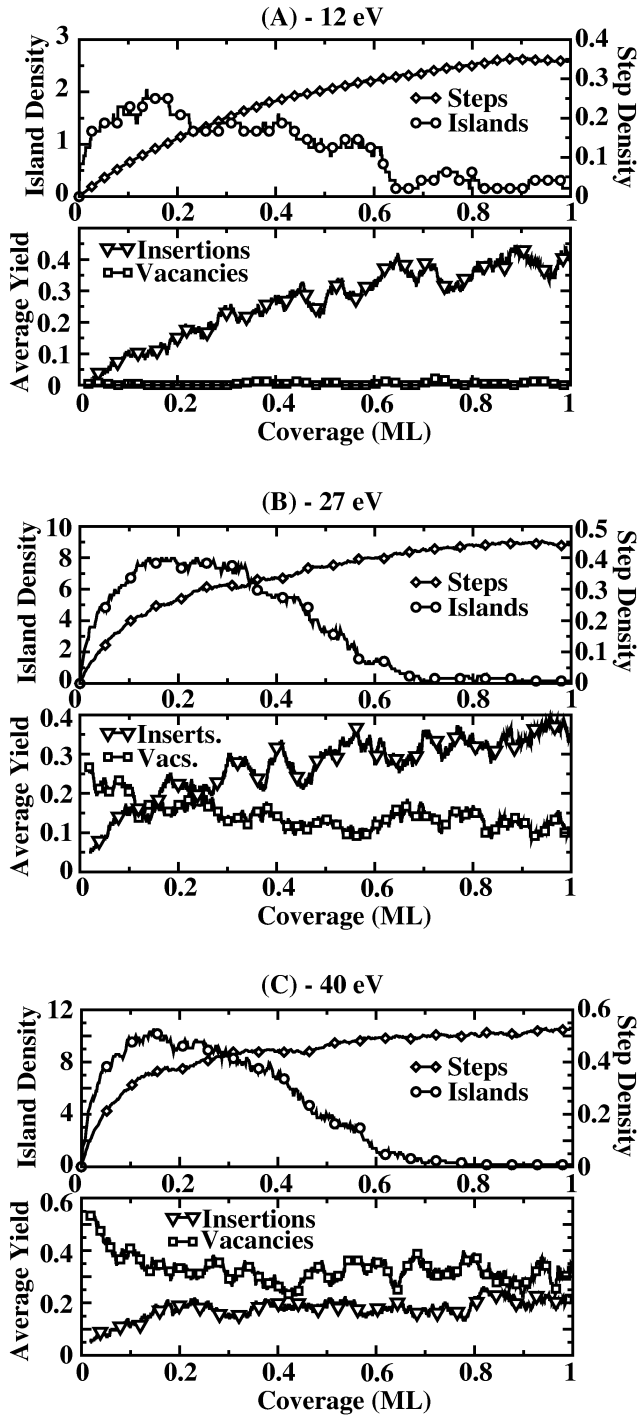}
\end{center}
\caption[]{The atomic configuration of the surface determines the
relative yields of mechanisms activated by the incident atom beam.
For each of three energies, 12 eV (A), 27 eV (B), and 40 eV (C), the
top panel show the island and step densities while the bottom panel
shows the time average yields for insertion and adatom-vacancy
production.  Beyond the first monolayer, these values reach
equilibrium and change very little about the 1 ML value.}
\label{fig:dyn-comp}
\end{figure}

The time evolution of the RMS roughness is shown in greater detail for
a few selected energies in Figure \ref{fig:roughvtime}.  The RMS
roughness for the thermally deposited film diverges as a power law, as
expected\cite{Fam}.  For all the energies shown, the RMS roughness
grows rapidly until about 0.5 ML.  Below 0.5 ML, films deposited with
energies greater than 20 eV actually develop roughness faster than the
thermally deposited film.  As the hyperthermal beam creates large
numbers of adatom vacancy pairs, the surface width increases rapidly,
but these extra adatoms in turn increase nuclei densities, which
contribute to higher step densities, shown in Figure
\ref{fig:dyn-comp}.

Island density and step density (top frames), as well as time averaged
insertion and vacancy yields (bottom frames) are presented in Figure
\ref{fig:dyn-comp}A-C for 12 eV, 27 eV, and 40 eV, respectively.
As the energy increases from 12 eV to 40 eV, the saturation island
density (approximately the density at 0.15 ML) increases by a factor
of six.  Since the island density and the average island size are
related ($N_x=\theta/\bar{s}$, where $\theta$ is the coverage and
$\bar{s}$ is the average island size), one might naively expect a
factor of six increase in island density to correspond with a
$\sqrt{6}$ increase in the step density at constant coverage.  In
actuality, the increase in island density does not translate directly
to an increase in step edge density, the step edge density in Figure
\ref{fig:dyn-comp} increases by only about 50\%, not $\sqrt{6}$.  This
is partly a consequence of our definition of island density.  We
define the step density as the number of atoms with empty neighbor
sites, since the insertion mechanism relies on displacing an atom into
an empty neighboring lattice position.  This definition can decouple
island density and step density.  For example, if all islands were
composed of four atoms, the step density would be 4 times the island
density.  But if all the islands were made from dimers, the step
density would be the same, but the island density would increase by a
factor of two.

Figure \ref{fig:dyn-comp} also displays the dynamic yields for the
insertion mechanism and the formation of vacancies at 12eV (A), 27 eV
(B), and 40 eV (C) respectively.  While the yields discussed in the
previous section and presented in Figure \ref{fig:mdresults} provide
an average yield for insertions and vacancies at a given energy near a
(111) step edge, the yields in Figure \ref{fig:dyn-comp} are dynamic.
Each line-shape is a running average a few hundreths of a monolayer in
coverage.  The yields for insertions and vacancy production is
observed to be sensitive to fluctuations in the surface structure.

 The insertion yields at 12 eV and 27 eV track the step density very
closely.  At very short times, while the surface is still very flat,
the step density is very low and few sites are available for
insertion.  During these times, the vacancy yield is large, which in
turn increases the island and step density.  With increasing step
density, more sites become available for insertion, increasing the
insertion yield.  At these energies, adatom-vacancy production is
suppressed at step edges.  So, as the surface becomes more populated
with islands and fewer flat terraces, the vacancy yield decreases.
At about 0.3 ML, islands begin to coalesce and decrease the number of
first layer step edges.  The formation of adatom-vacancy pairs on the
second layer (atop islands) keeps the step edge density high, and the
insertion mechanism does not suffer.

The increase in the island density and corresponding increase in the
step density below 0.5 ML in the 27 and 40 eV simulations sets the
stage for smooth growth at later times.  The drop in the island
density to a very low value by 0.7 ML indicates near completion of the
first layer before second layer growth.  The abrupt change in the RMS
roughness (Figure \ref{fig:roughvtime}) near 0.5 ML of coverage
illustrates the predicted benefit of using hyperthermal energy
particles.  With increasing particle energy the RMS roughness grows
more slowly until, by 40 eV, the RMS roughness does not perceptibly
increase above 1 ML of coverage.  This ``saturated'' roughness was
observed in all films grown with energies at or above 30 eV.  While
the roughness does not noticeably increase above 1 ML in these films,
higher incidence energies result in larger ``saturation'' roughnesses.

The effect of a saturated roughness is the result of the insertion
yield and the vacancy yield both achieving saturation.  At 27 eV, the
first four lattice positions immediately above a step edge are
available for insertion, which suppresses vacancy formation.  As a
result, insertions have a higher yield during growth above 0.5 ML than
vacancies (Figure \ref{fig:dyn-comp}B), and the saturation roughness
is less than at 40 eV.  At higher energies, the first lattice position
above the step is unstable upon impact.  As a result, the balance of
insertion and vacancy yields falls in favor of vacancies at 40 eV (see
Figure \ref{fig:dyn-comp}C, and notice the scale differences between
12 (A), 27 (B), and 40 eV (C)).

While total insertions decrease at 40 eV, other energetic effects
begin to compensate and keep the roughness from increasing
dramatically.  For example, the increasing atom energy can break
islands into smaller pieces, preventing an additional layer from
nucleating on top of it.  Atom impacts on top of multiple layers can
lead to collective downward mobility, e.g. at 40 eV two atoms or more
were observed to fall in the layer beneath the impact site one time in
twenty-five.  These and other mechanisms which involve collective
motion of multiple atoms have been discussed in detail
elsewhere\cite{Voter1,Wad1}.

The large insertion yield effectively reduces the interlayer diffusion
barrier by providing an alternative to thermal descent for crossing
the step.  We have performed KMC simulations using an reduced
interlayer diffusion barrier to mimic this effect.  This
oversimplification fails to reproduce the correct line-shape for the
RMS roughness and does not provide a layer-by-layer type growth,
underscoring the importance of adatom-vacancy pairs.  The
adatom-vacancy pairs contribute to establishing a microscopically
rough interface that sustains a macroscopically smooth growth front
through high insertion yields.

To review, as a film begins to grow with atoms in the 25 eV energy
range, the surface initially becomes pocked with vacancies.  The
deposited atoms combine with adatoms from adatom-vacancy pairs to
develop high densities of small islands, which have low probabilities
for second layer nucleation.  The vacancies and the new islands both
increase the step edge density, which leads to high insertion yields.
As the islands grow from insertion and aggregation, some begin to form
second layer islands and obtain vacancies prior to coalescence.  This
establishes an average distance between step edges of about three
atomic positions, and the roughness saturates.  As the vacancies are
filled and new levels are nucleated, the surface grows smoothly with a
constant roughness.  This smooth growth relies on both the insertions
to keep islands growing and the vacancies to provide additional
adatoms and to reduce the area available for new layer nucleation.

\section{Non-intuitive Role of Temperature and Flux in Energetic Deposition}
\label{sec:tempflux}

\begin{figure}[t]
\begin{center}
\epsfbox{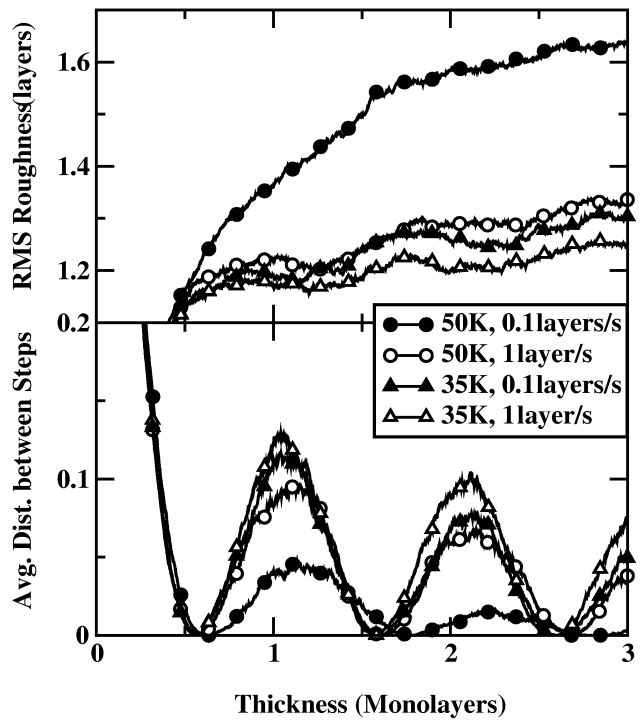}
\end{center}
\caption[]{In this re-entrant layer-by-layer mode, the roles of
temperature and flux have reversed behavior: high flux and low
temperatures yield the smoothest films, opposite thermal deposition.
Shown here are films grown at 0.1 ML/s and 1 ML/s flux, and at 35 K
and 50 K.  The top panel shows the RMS roughness of each of the four
films, and the bottom panel is the simulated anti-phase intensity.
Films grown at 35 K and 1 ML/s are 50 \% smoother than films grown at
50 K with 0.1 ML/s, with anti-phase intensity maxima more than three
times as intense.  The reversal of roles is a consequence of relying
on the insertion mechanism for smooth growth, which requires high step
densities to be effective.}
\label{fig:fluxtemp}
\end{figure}

During thermal homoepitaxy, the roles of temperature and flux are well
understood \cite{Rosen2}, smooth growth occurs when adatoms have
enough time to diffuse to an existing island and the islands have time
to coalesce before second layer nucleation occurs.  This is most
likely to occur when the temperature is increased to increase the
diffusion length, and the flux decreased to reduce the probability of
nucleating a new island before coalescence.

During hyperthermal energy deposition, this phenomenology reverses
due to the strong dependence on step density.  Other authors have
found the best results for smooth growth with hyperthermal deposition
can be obtained by maximizing the nuclei density\cite{Rosen1,Wulf1}.
Since island densities scale as $N_x \propto \left({F}/{D}\right)^p$
(F is flux, D is the temperature dependent diffusivity, and p depends
on the critical nuclei size)\cite{venables}, establishing a high
density of islands requires decreasing the temperature and increasing
the flux.  A high island density also means a small average island
size ($N_x={\theta}/{\bar{s}}$, where $\theta$ is the film coverage
and $\bar{s}$ is the average island size), reducing the target area
for second layer nucleation, and keeping the entire second layer
surface close to a step edge.  A hyperthermal atom incident on top one
of these islands has a very high probability of inserting, rather than
relying on kinetic diffusion to cross to the lower terrace.

We have presented RMS roughness as a function of time and anti-phase
intensity data for films grown with 24 eV atoms at various
temperatures and flux in Figure \ref{fig:fluxtemp}.  Contrary to
thermal deposition\cite{Rosen2}, the rougher films occur with the
higher temperatures and lower fluxes.  In addition, one can compensate
for a decrease in flux by decreasing the temperature.  For example,
the surface grown at 50K and 0.1 ML/s was much rougher than the film
grown at 50 K and 1 ML/s, but at 35 K with 0.1 ML/s flux, the film
grows smoothly.

\section{Hyperthermal Energy Induced Island Densities}
\label{sec:hypisls}

\begin{figure}[t]
\begin{center}
\epsfbox{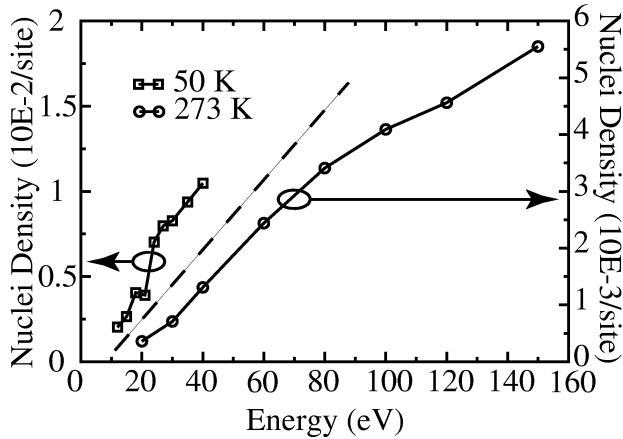}
\end{center}
\caption[]{With increasing energy, large yields of adatom-vacancy
pairs raise the free adatom density, which in turn increases the
island density.  This effect is observed both at 50 K, and at room
temperature, where the island density increases by more than an order
of magnitude with 150 eV of energy.  Plotted here is the maximum
island density achieved by 0.15 ML of coverage for 50 K deposition
(squares - left axis scale) at 1 ML/s and 273 K deposition (circles -
right axis scale) at 100 ML/s.}
\label{fig:Nxdens}
\end{figure}

While atomic insertion provides a compelling mechanism for controlling
surface roughness, the extremely small island sizes and high step
densities required for layer-by-layer behavior reduce the
effectiveness at temperatures typical for film growth.  At typical
deposition temperatures, the average island size becomes large enough
to reduce the number of sites available for insertion significantly,
in turn reducing the average insertion yield.  Even though the
insertion yield drops, the adatom-vacancy pairs generated by
hyperthermal energy ion beams provide additional adatoms that increase
island densities.

Island densities after the first layer reaches 0.15 ML of material are
shown for films grown at 50 K and 273 K in Figure \ref{fig:Nxdens}.
The 273 K deposition is performed on a 400x400 lattice at 100 ML/s
using all the diffusion moves listed in Tables \ref{tab:energies} and
\ref{tab:vacenergies}.   Increased adatom density due to
adatom-vacancy production leads to dramatic increases in the island
density at both temperatures.  The onset of sputter erosion near 40 eV
acts to slow the net growth speed relative to the nucleation rate,
increasing the effective flux.  As material in the substrate is lost
to sputter erosion, saturation island densities (typically at about
0.15 ML of coverage) are reached as low as 0.05 ML.  As atoms are
deposited or displaced, the first layer collects most of the material,
nucleating new islands while the additional vacancies produced by the
sputter erosion reduces the net meterial deposited.  For example,
consider an energy at which two adatom-vacancy pairs are produced and
the sputter yield is 0.5.  On average, every incident atom will create
2.5 atoms in the first layer, while only depositing 0.5 atoms total.
After 0.1 ML of {\em net} deposition, the first layer will have almost
0.25 ML coverage.

The data presented in Figure \ref{fig:Nxdens} is the density of
islands after the {\em first} layer has reached 0.15 ML of coverage,
not 0.15 ML of total deposition.  The saturation island density
depends on the coverage in the layer being considered, not on the
total amount of material deposited.  It may be surprising such a
strong effect remains at 273 K, as the thermal activation of diffusion
increases the probability for adatom-vacancy recombination.  We find
that the high adatom densities lead to rapid formation of dimers,
which still move freely with only a 117 meV diffusion barrier.  Dimer
step crossing is negligible, and dimer breakup occurs infrequently.

We have also found that high island densities are maintained with
unexpectedly low average island sizes due to islands being ``chipped''
by incident atoms.  While large islands are occasionally broken into
two smaller stable islands, breaking dimers and adatoms off of stable
islands occurs with a relatively high yield, on the order of 1/10
impacts on an island.  This contributes to additional nucleation of
small islands and suppresses the growth of large islands.

The significance of enhanced nuclei densities has already been
experimentally demonstrated by using an ion beam to increase the free
adatom density at the beginning of each monolayer of growth while
depositing copper on Cu(111)\cite{Wulf1}.  (In this example, a
separate 1.2 keV argon ion beam and a flux of thermal atoms were
used.)  The ability to dramatically increase nuclei densities using a
single growth beam in the room temperature regime opens new
possibilities for circumventing three-dimensional growth.  It is
possible that pulsing the final beam energy between a low energy
($\sim$20 eV) and a high energy ($\sim$100 eV) in manner similar to
the previous example\cite{Wulf1} can allow the benefits of high island
density and atomic insertion to be utilized while using a single
deposition source.

\section{Acknowledgements} 
The authors would like to acknowledge Oana Malis, Markus Rauscher, and
Joel Brock for their discussions and editorial contributions.  This
work was supported by the Cornell Center for Materials Research
(CCMR), a Materials Research Science and Engineering Center of the
National Science Foundation (DMR-0079992). Portions of this research
were conducted using the resources of the Cornell Theory Center, which
receives funding from Cornell University, New York State, Federal
Agencies, and corporate partners.  Additional support was provided by
the AFOSR under grant number F49620-97-1-0020.

\bibliographystyle{prsty}
\bibliography{submitted}

\end{document}